\documentstyle[aps,12pt,tighten,amssymb,axodraw,cite]{revtex}

\title{\vspace*{1.5in}D-Brane Recoil and 
Supersymmetry Breaking as a Relaxation Process}

\author{ {\bf A. Campbell--Smith}$^{a}$ and 
{\bf N.E.~Mavromatos}$^{a,b}$ }

\address{${}^a$Theoretical Physics (University of Oxford), 1 Keble
  Road, OXFORD, OX1 3NP, U.K.,\\${}^b$Theory Division, CERN, CH-1211
  Geneva 23, Switzerland.}

\begin{document}

\maketitle

\vspace*{1cm}
\begin{centering} 

{\bf Abstract} 

\end{centering}
\begin{abstract}
We propose a new mechanism for the formation of conical singularities
on D-branes by means of recoil resulting from scattering of closed
string states propagating in the (large) transverse dimensions.  By
viewing the (spatial part of the) 
four-dimensional world as a 3-brane with large
transverse dimensions the above mechanism can lead to supersymmetry
obstruction at the TeV scale.  The vacuum remains supersymmetric while
the mass spectrum picks up a supersymmetry obstructing mass splitting.
The state with ``broken'' supersymmetry is not an equilibrium ground
state, but is rather an excited state of the D-brane which relaxes to
the supersymmetric ground state asymptotically in (cosmic) time.
\end{abstract}

\vspace*{-5in}
\begin{flushright}
  OUTP-99-36P\\
  CERN-TH/99-250\\
  hep-th/9908139
\end{flushright}
\vspace*{5in}

\newpage
The issue of supersymmetry breaking remains unresolved; while
supersymmetry provides a natural explanation for the vanishing of the
cosmological constant (vacuum energy) and offers a resolution of the
hierarchy problem it is not a symmetry of the low-energy world.
Softly broken supersymmetry can still control the Higgs mass and
resolve the hierarchy problem, but radiative corrections to Standard
Model processes measured in precision electroweak experiments at LEP
coupled with direct Higgs and sparticle searches require that the
scale of supersymmetry breaking (i.e. the masses of the lightest
sparticles) be no more than a few TeV without necessitating unnatural
fine tuning of the Standard Model parameters~\cite{chankowski98}.
Some years ago a novel scenario for generating massive sparticles
while maintaining the vanishing cosmological constant was
proposed~\cite{witten95}.  The scenario referred to
\((2\!+\!1)\)-dimensional supergravity theories and, instead of
breaking supersymmetry, involved the {\em obstruction} of
supersymmetry by massive states in the spectrum of supergravity living
in a spacetime with conical singularities.  In fact, as pointed out in
reference~\cite{witten95} based on work in reference~\cite{deser84},
any state in \((2\!+\!1)\) dimensions which has non-zero energy
produces a geometry which is asymptotically conical.  In such
spacetimes there are no covariantly constant spinors; given that in
supersymmetry the unbroken supercharges are spinor fields which are
covariantly constant at infinity, this suggests that there are no
unbroken supersymmetries in \((2\!+\!1)\)-dimensional supergravity
theories.  The Bose--Fermi degeneracy in the massive spectrum is
lifted in proportion to the deficit angle \(\delta\) of the conical
geometry,
\begin{equation}
\delta = 2\pi G_3 \delta m
\label{split}
\end{equation}
where $m$ is the mass splitting, and $G_3$ is Newton's constant in
three dimensions~\cite{deser84}.  However, the vacuum energy
(cosmological constant) remains zero, given that the vacuum state
remains supersymmetric.

An explicit realization of the above phenomenon in the context of a
specific supergravity model in three spacetime dimensions has been
provided in reference \cite{becker95}, while conical singularities and
supersymmetry obstruction in the context of ${\mathcal N}\!=\!1$
supergravity in four spacetime dimensions have been discussed in
\cite{nishino96}.  The presence of conical singularities on any
Einstein manifold $X$ will in general lead to a complete breaking of
supersymmetry, except for special choices of the manifold $X$, in
which case some supersymmetries may survive.  Their number depends on
the number of covariantly constant spinors (or equivalently the number
of Killing spinors) on the manifold $X$. For certain geometries a
classification of the unbroken supersymmetries is
complete~\cite{bar93}..  In the present work we shall be interested in
the case where no supersymmetries are left unobstructed; in particular
we shall be interested in four-dimensional ${\mathcal N}\!=\!1$
supergravity models, viewed as low-energy field theories of some
string (or D-brane) theories.

The point of this article is to describe what is in our opinion a
novel way to generate conical singularities on the four-dimensional
world by adopting the modern
view~\cite{dimopo98,antoniadis98,antoniadis99} that the
four-dimensional spacetime we observe is actually a D-3-brane living
in a higher-(ten or eleven)-dimensional universe.  Only closed string
states (gravitons) propagate in the dimensions transverse to the brane
while gauge and matter fields are described by open string states
ending on the brane.  Consistent embeddings of this idea in detailed
string models have recently been discussed in
references~\cite{antoniadis98,antoniadis99}.

Despite the fact that in such a picture the detailed dynamics of the
bulk higher-dimensional spacetime is not fully known, several
non-trivial predictions for the (low-energy) physics on
four-dimensional spacetime can emerge~\cite{dimopo98}.  In this
article we shall point out yet another prediction, that of possible
supersymmetry obstruction at the TeV scale.

We first note an important ingredient of closed-string scattering:
that of the resulting recoil of the
D-brane~\cite{kogan96,ellis96,kanti98,mavro+szabo}.  This has been
ignored in most discussions of the scattering process, such as
emission of closed string states into the bulk and/or absorption by
the brane.  From a worldsheet viewpoint the recoil is
described~\cite{kogan96} by deformations of the pertinent
\(\sigma\)-model which obey a logarithmic conformal
algebra~\cite{lcft}.  Field theories described by such algebras lie on
the border between conformal field theories and general renormalizable
two-dimensional quantum field theories.  The fact that recoil of a
D-brane is described not by an ordinary conformal field theory but by
a logarithmic one is associated with the fact that the recoil process
describes a change of state in the $\sigma$-model background, and as
such is a non-equilibrium process.  This is
reflected~\cite{ellis96,mavro+szabo} in the logarithmic operator
algebra itself.

It has been argued recently~\cite{adrian99:3} that when properly taken
into account the recoil may lead to non-trivial phenomena on the
brane such as stochastic fluctuations in the arrival times of photons
propagating on the brane.  Given that such effects are considerably
larger in size than the string scale, it is evident that they are
further enhanced in the picture of reference~\cite{dimopo98} where the
string scale in the bulk is of the order of a TeV.  In
reference~\cite{adrian99:3} this has been used to place bounds on
consistent string models of large extra
dimensions~\cite{antoniadis99}.

In the present article the recoil process will be discussed in
conjunction with another phenomenon that characterizes such theories,
namely supersymmetry obstruction on the brane.  Our objective is to
discuss the appearance and nature of the conical singularities due to
the recoil process, mentioned in references \cite{kogan96,adrian99:3},
and then to estimate the order of magnitude of the induced
supersymmetry obstruction~\cite{witten95}.

As discussed in references~\cite{kogan96,ellis96,mavro+szabo} in the
case of D-brane string solitons, their recoil after interaction with
a closed string (graviton) state is characterized in a worldsheet
context by a $\sigma$-model deformed by pairs of logarithmic
operators~\cite{lcft}:
\begin{equation}
C^I_\epsilon \sim \epsilon \Theta_\epsilon (X^I),\qquad D^I_\epsilon
\sim X^I \Theta_\epsilon (X^I), \qquad I \in \{0,\dots, 3\}
\label{logpair}
\end{equation} 
defined on the boundary $\partial \Sigma$ of the string
worldsheet. Here $X^I$ obey Neumann boundary conditions on the string
worldsheet, and denote the brane coordinates.  The remaining $y^i,
i\in \{4, \dots, 9\}$ denote the transverse directions.  In the case
of D-particles, examined in
references~\cite{kogan96,ellis96,kanti98,mavro+szabo}, $I$ takes the
value $0$ only, in which case the operators (\ref{logpair}) act as
deformations of the conformal field theory on the worldsheet: the
operator \[U_i \int _{\partial \Sigma} \partial_n X^i D_\epsilon\]
describes the shift of the D-brane induced by the scattering, where
$U_i$ is its recoil velocity, and \[Y_i \int _{\partial \Sigma}
\partial_n X^i C_\epsilon\] describes quantum fluctuations in the
initial position $Y_i$ of the D-particle. It has been
shown~\cite{mavro+szabo} that energy-momentum is conserved during the
recoil process.  We also note that $U_i = g_s P_i$, where $P_i$ is the
momentum and $g_s$ is the string coupling, which is assumed here to be
weak enough to ensure that D-branes are very massive, with mass
$M_D=1/(\ell _s g_s)$, where $\ell _s$ is the string length. From
these one obtains $U_i = \ell_s g_s ( k^1_i + k^2_i)$, where $k^1\
(k^2)$ is the momentum of the propagating closed string state before
(after) the recoil~\cite{mavro+szabo}.

\begin{figure}
\begin{center}
\begin{picture}(100,155)(0,-5)
\Oval(28,78)(20,10)(0)
\CBox(28,58)(38,98){White}{White}
\Line(28,58)(50,78)
\Line(28,98)(50,78)
\Vertex(49.8,78){.3}
\Line(10,10)(10,110)
\Line(10,10)(45,45)
\Line(10,110)(45,145)
\Line(45,45)(45,73.5)
\Line(45,82.5)(45,145)
\Text(12,33)[l]{$D_1$}
\PhotonArc(75,78)(12,0,360){1.7}{9}
\LongArrow(80,78)(100,78)
\DashLine(80,78)(63,78){2}
\Line(63,78)(54,78)
\DashArrowLine(28,33)(28,60){3}
\DashArrowLine(28,96)(28,123){3}
\DashArrowArcn(74.5,78)(50,200,160){3}
\end{picture}
\caption{Schematic representation of the recoil effect: the surface of
the D-brane D1 is distorted by the conical singularity that results
from closed string emission into the bulk. The dashed line on D1
represents the (disturbed) trajectory of a matter particle living on
the brane. \label{fig1}}
\end{center}
\end{figure}
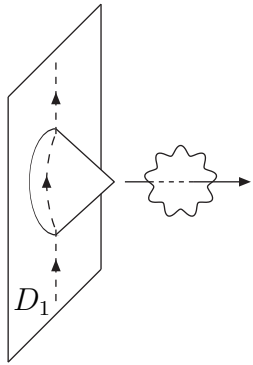

In the case of D-$p$-branes, the pertinent deformations are slightly
more complicated. As discussed in reference~\cite{kogan96}, the
deformations are given by \[\sum_{I} g^1_{iI} \int _{\partial \Sigma}
\partial_n X^i D_\epsilon ^I \qquad{\mathrm and}\qquad\sum_{I}
g^2_{Ii} \int _{\partial \Sigma} \partial_n X^i C^I_\epsilon. \] The
$0i$ component of the ``tensor'' couplings
$g^{\alpha}_{Ii},~\alpha\in\{1,2\}$ include the collective momenta and
coordinates of the D-brane as in the D-particle case above, but now
there are additional couplings $g^{\alpha}_{Ii},~I\ne0$, describing
the ``folding'' of the D-brane under the emission of a closed string
state propagating in a transverse direction, as shown schematically in
Fig.~\ref{fig1}.  Intuitively it is clear that this emission and the
resulting recoil results in a wedge-shaped ``folded'' conical space,
like a surface tension effect on the higher-dimensional analogue of an
elastic membrane.  In the following we will verify this for the
D-particle case after correctly Liouville-dressing the deformation
operators.

The correct specification of the logarithmic pair in equation
(\ref{logpair}) entails a regulating parameter
$\epsilon\rightarrow0^+$, which appears inside the
$\epsilon$-regularized $\Theta_\epsilon (t)$ operator:
\[\Theta_\epsilon (X^I) = \int \frac{d\omega}{2\pi}\frac{1}{\omega
-i\epsilon} e^{i\omega X^I} .\] In order to realize the logarithmic
algebra between the operators $C$ and $D$, one takes~\cite{kogan96}:
\begin{equation} 
\epsilon^{-2} \sim \ln [L/a] \equiv \Lambda,
\label{defeps}
\end{equation} 
where $L$ ($a$) are infrared (ultraviolet) worldsheet cutoffs.  The
recoil operators (\ref{logpair}) are slightly relevant, in the sense
of the renormalization group for the worldsheet field theory, having
small conformal dimensions $\Delta _\epsilon = -\epsilon^2/2$.
Thus the $\sigma$-model perturbed by these operators is not conformal
for $\epsilon \neq 0$, and the theory requires Liouville
dressing~\cite{david88,distler89,ellis96,kanti98}.

To determine the effect of such dressing on the spacetime geometry,
it is essential to write~\cite{ellis96} the boundary recoil
deformations as bulk worldsheet deformations by partial integration
\begin{equation}
    \int _{\partial \Sigma} g^1_{iI} X^I\Theta_\epsilon (X^I) 
\partial_n X^i =
\int _\Sigma \partial_\alpha \left(g^1_{iI} X^I\Theta_\epsilon (X^I) 
\partial ^\alpha X^i \right) 
\label{a1}
\end{equation}
These operators can be made marginal on a curved worldsheet by
Liouville-dressing~\cite{distler89}.  One Liouville-dresses the (bulk)
integrand by multiplying it by a factor $e^{\alpha_{Ii}\phi}$, where
$\phi$ is the Liouville field and $\alpha_{Ii}$ is the gravitational
conformal dimension, which is related to the flat worldsheet anomalous
dimension $-\epsilon^2/2$ by
\begin{equation}
\alpha_{Ii}=-\frac{Q_b}{2} + 
\sqrt{\frac {Q_b^2}{4} + \frac {\epsilon^2}{2} } 
\label{anom}
\end{equation}
where $Q_b$ is the central-charge deficit of the bulk worldsheet
theory.  In the recoil problem at hand, as discussed
inreference~\cite{kanti98}, $Q_b^2 \sim \epsilon^4/g_s^2$ for weak
folding deformations $g_{Ii}^1$.  This yields $\alpha_{Ii} \sim
-\epsilon $ to leading order in perturbation theory in $\epsilon$, to
which we restrict ourselves here.

We next remark that, as the analysis of reference~\cite{ellis96}
indicates, the $X^I$-dependent field operators $\Theta_\epsilon (X^I)$
scale as follows with $\epsilon$: $\Theta_\epsilon(X^I) \sim
e^{-\epsilon X^I} \Theta(X^I)$, where $\Theta(X^I)$ is the normal (not
$\epsilon$-regularized) Heaviside step function without any field
content, evaluated in the limit $\epsilon \rightarrow 0^+$.  The bulk
deformations, therefore, yield the following $\sigma$-model terms:
\begin{equation} 
\epsilon  g^1_{iI} X^I 
e^{\epsilon(\phi_{(0)} - X^I_{(0)})}\Theta(X^I_{(0)}) \int _\Sigma
\partial^\alpha X^I \partial^\alpha X^i
\label{bulksigma}
\end{equation} 
where the subscripts $(0)$ denote worldsheet zero modes.  Upon the
interpretation of the Liouville zero mode $\phi_{(0)}$ as target time
$t$, the deformations (\ref{bulksigma}) yield spacetime metric
deformations in a $\sigma$-model sense, which were interpreted
in reference~\cite{ellis96} as expressing the distortion of the spacetime
surrounding the recoiling D-brane soliton.

We choose to work in a region of spacetime on the D-3-brane such that
$\epsilon (\phi - X^I)$ is finite in the limit $\epsilon \rightarrow
0^+$.  The resulting spacetime distortion is therefore described by
the metric elements
\begin{equation} 
G_{0i} = \epsilon g^1_{iI} X^I \Theta (X^I) +{\mathcal O}(\epsilon^2).
\label{gemteric}
\end{equation}  
The presence of the $\Theta(X^I)$ functions indicate that the induced
spacetime is piecewise continuous~\footnote{The important
implications for non-thermal particle production and decoherence for a
spectator low-energy field theory in such spacetimes were discussed
in~\cite{kanti98,ellis96}, where only the D-particle recoil case
was considered.}.

To see how the Liouville dressing above leads to conical spacetimes,
it is sufficient to restrict attention to the D-particle
case~\cite{kanti98}.  In that case, the resulting spacetime resembles
flat Minkowski spacetime to ${\cal {O}}(\epsilon
^{2})$~\cite{kanti98}, upon making the following transformation for $t
> 0$:
\begin{equation} 
{\tilde X^i} = X^i + \frac{1}{2}\epsilon U^i t^2, \qquad  
{\tilde t}=t .
\label{rindler}
\end{equation} 
This implies that the spacetime induced by the recoiling D-brane
resembles, for $t \gg 0$ and to order ${\cal O}(\epsilon^{2})$, a
Rindler wedge space with (non-uniform~\cite{kanti98}) `acceleration'
$\epsilon U^i$. This space is well known to produce a conical
singularity when the (Euclidean) time is compactified over an inverse
`temperature' interval, and has bulk deficit angle
\begin{equation}
\delta_{0i} \sim 2\pi \left(1 - 1/\epsilon |U^i| \right)~.
\label{rindler2}
\end{equation} 
Such conical singularities lead to supersymmetry breaking in the bulk,
not on the D-particle's world-line.  An interesting realization of
this D-particle case is that of a D-0-brane defect embedded in a
D-$p$-brane, i.e. dimensionally reduced intersecting D-branes.  In the
deficit above the r\^{o}le of the `temperature' is played by the
`acceleration' $\epsilon |U^i|$~\cite{unruh76}.  Indeed, the conical
spaces described here have a similar effect to non-zero temperatures
which can be computed using a thermal superspace
formalism~\cite{derendinger98}: there are thermal \emph{mass
splittings} between superpartners, proportional to $T$ (for particles
massless at $T=0$ like the graviton and photon and their
supersymmetric partners).  Our discussion above refers, rather
generically, to mass splittings on the D-$p$-brane (specifically a
D-3-brane).  However, for scattering events like the one depicted in
Figure~\ref{fig1}, the formalism also implies similar ``thermal'' mass
splittings in the bulk.  Let the compactification volume (in units of
the (bulk) string scale $M_s^{-1}$) of the extra $n$ dimensions be
denoted by $\Omega = (\Lambda M_s)^n$, where $\Lambda$ is the radius
of the extra dimensions (assumed, for simplicity, to be of equal
size).  The scale $\Lambda$ is related to the Planck scale $M_P$ on
the D-$p$-brane by~\cite{dimopo98}: $M_P = M_s/g_s = \left( \Lambda
M_s \right)^{n/2}.$ Then a (``thermal'') mass splitting on the
D-3-brane, $\delta M_{\mathrm{D3}}$, can be estimated by naive
dimensional reduction from the corresponding one in the bulk geometry,
$\delta M_b \sim M_s \epsilon |U_i|$, as:
\[\delta M_{\mathrm{D3}} \sim \delta M_b \Omega = M_P \left(1/g_s \right) 
\epsilon |U_i|.\]
Taking into account energy-momentum conservation during the recoil
process~\cite{mavro+szabo} one has: $U_i=g_s \Delta P_i/M_s$, where
$\Delta P_i$ is a typical momentum transfer in the bulk, along the
direction of recoil.  In such a case,
\[\delta M_{\mathrm{D3}} \sim M_P \,\epsilon \left(\Delta P_i/M_s \right).\]
As we shall discuss below, if the scattering occured at very early
times, then a typical bulk scattering energy would be of order $M_s$,
which implies that the induced supersymmetry-obstructing mass
splittings on the D-3-brane would be of order $\epsilon M_P$.

When the above formalism is extended to the full recoiling D-$p$-brane
case the deformations arising from localized emissions (e.g. closed
strings or heavy D-particles) into the bulk induce a wedged or conical
(for the deformation should be symmetric around the axis of recoil)
world-volume for the brane also; see Figure~\ref{fig1}.  This excited
state of the brane is expected to lead to mass-splittings proportional
to the properly renormalized recoil couplings $g_{iI}$, being proper
generalizations of the D-particle operators $Y^i$ and $U^i$.  In the
following we will show how careful interpretation of the
Liouville-dressing, and proper identification of the parameter
$\epsilon$ as the physical time, lead to the possibility of
supersymmetry obstruction well below the natural scale $M_P$.

To this end, we recall that the worldsheet two-point correlation
functions of the recoil operators have the following
form~\cite{kogan96}:
\begin{eqnarray}
&~&\langle C_\epsilon (z)C_\epsilon (0)\rangle \stackrel{\epsilon\to
0}{\sim} 0+ {\cal O}(\epsilon^2) \nonumber \\ &~&\langle C_\epsilon
(z)D_\epsilon (0)\rangle \stackrel{\epsilon\to 0}{\sim}
{\pi\over2}\sqrt{{\pi\over\epsilon^2\Lambda}} \left(1+2\epsilon^2
\log|z/a|^2 \right) \nonumber \\ &~&\langle D_\epsilon (z)D_\epsilon
(0)\rangle =\frac{1}{\epsilon^2} \langle C_\epsilon (z)D_\epsilon
(0)\rangle \stackrel{\epsilon\to 0}{\sim}
{\pi\over2}\sqrt{{\pi\over\epsilon^2\Lambda}}
\left({1\over\epsilon^2}+2\log|z/a|^2\right)
\label{twopoint}
\end{eqnarray} 
which in the limit $\epsilon \rightarrow 0^+$ gives the logarithmic
algebra~\cite{lcft} modulo the leading divergence in the $\langle
D_\epsilon D_\epsilon\rangle$ recoil correlation function.

The identification (\ref{defeps}) turns out to be very important for
our purposes here.  As discussed in references~\cite{kogan96,lizzi97},
under worldsheet scaling transformations parametrized by variations
of the cutoff \[L\longmapsto L'=Le^t \qquad \Rightarrow \qquad \epsilon^2
\longmapsto {\epsilon'}^2 = \frac{\epsilon^2}{1+2\epsilon^2 t},\] then as
a result of the logarithmic algebra (\ref{twopoint}) the operators $C$
and $D$ transform as
\begin{eqnarray}
D_\epsilon &\longmapsto& D_{\epsilon '} = D_\epsilon + t
C_\epsilon,\nonumber\\
C_\epsilon &\longmapsto& C_{\epsilon '} = C_\epsilon\nonumber
\end{eqnarray}
which implies a similar transformation for the couplings.  In
particular, the the $g^2_{Ii}$ bending couplings are shifted as
\[g^2_{Ii} \longmapsto (g^2_{Ii})' = g^2_{Ii} +
g_{Ii}^1 t,\] while the $g_{Ii}^1$ couplings remain invariant.  From
this and the fact that $g_{0i}^{2,1}$ are interpreted respectively as
the collective coordinates and momenta of the recoiling D-brane, one
observes that the scale $\epsilon^{-2}$ may be interpreted as a
Galilean time for the (heavy) defect system.

It is important to understand the connexion of this time with the
physical time as measured by an observer on the brane.  To answer this
question we remark that the worldsheet renormalization group scale
$\ln|L/a|^2$ may be associated with the zero-mode of the Liouville
field~\cite{david88,distler89}, which in turn is identified with the
target time $t$ on the brane, as justified in detail in
references~\cite{ellis96,mavro+szabo}.  From this one immediately has the
identification \[\epsilon^{-2} \longleftrightarrow \eta t,\] where the
time $t$ is measured in string units $t_s\!=\!\ell_s/c$, where $\ell_s$ is
the string length.  The constant of proportionality $\eta$ can be
determined as follows:  the Liouville field used to dress and hence
restore conformal invariance~\cite{david88,distler89} in the
non-conformal $\sigma$-model perturbed by the recoil
operators~\cite{ellis96} has a kinetic term in the $\sigma$-model of
the form:
\begin{equation}
\int_{\Sigma} d^2z \; Q^2 (\partial \phi)^2 
\label{liouvkin}
\end{equation}  
where $\Sigma$ denotes a (closed string) worldsheet surface.  The
central charge deficit $Q^2=C[g]-C^*$ is written in terms of the
running central charge $C[g]$ given by the Zamolodchikov $C$-function
which can in turn be defined by an appropriate combination of the
two-point worldsheet correlation functions of the stress tensor for
the $\sigma$-model \(\langle T_{\alpha\beta}T_{\gamma\delta}\rangle\)
(the indices run over worldsheet coordinates).  For closed string
excitations the worldsheet is assumed to be a sphere, with Euler
characteristic $\chi\!=\!2$.  This implies that the worldsheet
correlation functions entering in the expression for the $C$-function
will have a prefactor of $1/g_s^2$.  For the weakly coupled string
theories in which we are interested, and for which the recoil
formalism of references~\cite{kogan96,mavro+szabo} applies, the
detailed analysis of reference~\cite{mavro+szabo} demonstrated that
the identification of the Liouville mode $\phi$ with the target time
$t$ leads to a consistent interpretation of the central charge deficit
$Q^2$ in the deformed $\sigma$-model. 
The recoil is described by 
an effective Lagrangian in target space of Born--Infeld type 
as a result of the identification:
\[Q^2 \sim \frac{1}{g^2_s}\sqrt{1 - |g_{Ii}^1|^2 + \dots}\]
to leading orders in worldsheet perturbation theory. 

The set of bending couplings $g_{Ii}^1 \equiv g_{Ii},~I\in\{0, \dots, p\},
~i\in\{p+1, \dots, 9\}$, are
relevant couplings with a worldsheet renormalization-group
$\beta$-function of the form
\begin{equation} 
  \beta_{g} = \frac{d}{d t} g_{Ii} = -\frac{1}{2t} g_{Ii} , \qquad 
t \sim \epsilon ^{-2} 
\label{betaf}
\end{equation} 
which implies that one may construct an exactly marginal set of
couplings ${\overline g_{Ii}}$ by redefining
\begin{equation}
{\overline g_{Ii}} \equiv \frac{g_{Ii}}{\epsilon}
\label{marginal}
\end{equation}
The renormalized couplings ${\overline g_{0i}}$ in \cite{mavro+szabo} play the
r\^ole of the physical recoil velocity of the D-brane, while the remaining
${\overline g_{Ii}},~I\ne 0$, describe the bending of the D-$p$-brane, $p\ne 0$.

By following an analysis similar to that for the D-particle case in
reference~\cite{mavro+szabo} it can easily be shown that the
renormalized bending couplings ${\overline g}_{Ii}$ are related to the
sum of momenta of the closed string states along the transverse
directions $k^{1,2}_i$ as follows:
\begin{equation} 
 {\overline g_{Ii}} \sim \frac{g_s}{M_s} (k^1_i + k^2_i), \qquad I\in\{0,
 \dots, p\},  \qquad i\in\{p+1, \dots, 9\}\label{bendingcoupl}
\end{equation} 
where $M_s \sim \ell _s^{-1}$ is the string scale (in units where 
$\hbar\!=\!c\!=\!1$). 

In this way we find that ${\overline g_{Ii}}\!=\!{\mathcal O}(1)$ for
closed string excitations with Planckian energies of order $M_s/g_s$,
in which case $Q^2 \rightarrow 0$. For any other low-energy state $E
\ll M_s/g_s$ $Q^2 \sim 1/g_s^2$.  We now notice that from a target
spacetime point of view such a kinetic term contributes to the
$G_{00}$ temporal component of the metric.  In order to obtain a
Robertson--Walker type of target-space universe under the above
identifications, it is crucial that we rescale the Liouville mode
$\phi$ to $Q\phi$ before identifying it with the (observable) cosmic
time $t_{\rm phys}$. In this way, from (\ref{defeps}), one obtains
\begin{equation} 
     \epsilon^{-2} = g_s t_{\rm phys}
 \label{phystime}
\end{equation}
where $t_{\rm phys}$ is the physical (cosmic) time in string units
$t_s$, pertaining to a Robertson--Walker universe.

The identification (\ref{phystime}) implies that the recoil/bending
deformation operators $g_{Ii}\int _{\partial \Sigma} D$ can be
expressed in terms of the marginal couplings ${\overline g_{Ii}}$
through (reinstating the string timescale $t_s$)
\begin{equation} 
g_{Ii} \rightarrow \frac{g_s}{M_s} (k^1_i + k^2_i)\left(\frac{t_s}{g_s
t_{\mathrm phys}}\right)^{1/2}
\label{physical}
\end{equation}
which in turn implies that the conical singularities on the D-brane
due to these deformations will have a deficit which will be decaying
with time as $t_{\mathrm phys}^{-1/2}$, as the system relaxes towards
its ground state.  Therefore if we view the world as a D-brane living
in a higher-dimensional string (or M-theory) universe, then a
scattering process whereby a closed string state propagates in the
transverse extra dimensions will excite the D-brane through recoil.
This will create a conical singularity at the time of the scattering
event whose formation can be described by deforming the worldsheet
$\sigma$-model by recoil operators.  It is crucial to identify the
worldsheet renormalization-group scale with the target time for the
mathematical consistency of the logarithmic
algebra~\cite{ellis96,mavro+szabo}. This identification naturally
implies a relaxation process for the recoiling brane.

During this relaxation process the presence of a conical singularity
with a deficit implies supersymmetry obstruction with mass splittings
given by a formula analogous to (\ref{split}), where now $G$ should be
Newton's constant on the D-brane.  The latter is related to the
four-dimensional Planck mass scale $M_P^{(4)} \sim M_s /g_s \sim
10^{19}$ GeV.  Thus the induced supersymmetry-obstructing mass
splitting $\delta m$ would be of the form:
\begin{equation} 
\delta m \sim g_{Ii} M_P^{(4)} \sim M_P^{(4)} g_s \frac{\left|k^1_i +
k^2_i\right|}{M_s} \left(\frac{t_s}{g_s t_{\mathrm phys}}\right)^{1/2}
\label{split2}
\end{equation} 
The result (\ref{split2}) has to be interpreted with care.  First of
all one should note that asymptotically in time $t_{\mathrm phys} \rightarrow
\infty$ the splitting tends to zero, implying the restoration of
supersymmetry in the spectrum.  This is natural from the point of view
adopted here, i.e. that the world we live on is a 3-brane in a
recoiling excited state after scattering with a closed string state.
The low-energy supersymmetry obstruction today is a result of the
relaxation of the brane.

In this scenario, at early times $t_{\mathrm phys} \sim t_s$, the
D-brane had a small size, of order of the string length $\ell _s$.
Therefore the closed string states trapped on it should have
uncertainties in energies of order $M_P^{(4)}$, implying that at times
$t_{\mathrm phys} \sim t_s$ after the initial scattering event, the
recoiling D-brane would have experienced obstructed supersymmetry
with mass splittings
\begin{equation} 
\delta m (t_s) \sim M_P^{(4)}   
\label{planckian}
\end{equation} 
This initial splitting diminishes as the time elapses according to
\begin{equation}
\delta m \sim M_P^{(4)} \left(\frac{t_s}{g_s t_{\rm
phys}}\right)^{1/2},
\label{split3}
\end{equation} 
One may arrange for the present day supersymmetry obstruction to be of
order a few TeV by selecting appropriately the ``frequency'' of the
scattering events, i.e. the quantities $t_{\rm phys}$, $g_s$ and
$t_s$.  For instance, for a scattering event occurring at early
cosmological times, e.g. at the time of last scattering $t_{\mathrm
phys} \sim 10^{26}/c$, for ``large'' string sizes $t_s \simeq
10^{-27}$~s, as in the scenario of reference~\cite{dimopo98}, and
small couplings $g_s \simeq 10^{-14}$, as required for a consistent
string theory embedding of such scenarios in type IIB closed string
theories~\cite{antoniadis99}, one obtains from (\ref{split3}) a
supersymmetry obstruction scale today of order of a few TeV.

In the type I$'$ open string case~\cite{antoniadis98} there exist
D-3-branes as solutions in the model, with the restriction that only
gravitational closed string states can propagate in the bulk, exactly
as required by the picture of reference~\cite{dimopo98}.  In this
model the string coupling $g_s$ is given by the Yang--Mills fine
structure constant at the string scale,
\[g_s = 4 \alpha_G (M_s).\]  According to reference~\cite{adrian99:3}
this results in stochastic fluctuations in the arrival times of
photons with energy $E$ and travelling a distance $L$ of
\[\Delta t \sim \alpha_G \frac{LE}{M_s}.\]  Astrophysical data on
gamma ray bursters are sensitive~\cite{amelino98} to $\Delta t \sim
LE/M_{\rm QG}$ with $M_{\rm QG}\sim 10^{15}$~GeV, whence the type I$'$
model seems incompatible with a low (TeV) string
scale~\cite{adrian99:3}, and hence also with TeV scale supersymmetry
obstruction by the mechanism described here.

In this article we have presented a mechanism by which supersymmetry
can be obstructed on our world at the TeV scale as a result of
D-brane recoil within the large extra dimension picture of
reference~\cite{dimopo98}.  Supersymmetry is obstructed by a Planck
scale mass splitting on the brane as a result of the formation of a
conical singularity.  As this excited state of the brane relaxes back
to the ground state the scale of supersymmetry obstruction is lowered.
We stress that this is a non-equilibrium process, and that the
relaxation we have described differs fundamentally from the ``slow
rolling'' by which a false vacuum decays to a true vacuum.  In our
case supersymmetry is {\em obstructed}~\cite{witten95} so that the
vacuum remains supersymmetric, and hence 
the cosmological constant on the brane vanishes, 
and it is only the matter spectrum which
does not respect supersymmetry as a result of the excited state of the
brane.

As stressed above, 
this scenario for supersymmetry obstruction 
is based on the Liouville 
(non-critical-string)
approach to D-brane recoil,
which involves the identification of the target time with the Liouville 
mode. 
Moreover, 
for the scenario to yield viable phenomenological 
predictions, it is essential that the 
extra dimensions are relatively ``large'', compared with the Planck scale, 
and that the emission of closed string 
states from the brane into the bulk is a very rare event. 
Such rare emissions, although 
compatible with the depletion of closed string states 
due to inflation, are still far from being 
understood at a satisfactory level of mathematical 
rigour in the context of the Liouville-string approach 
to D-brane recoil. This would  
require a 
detailed knowledge of 
the udnerlying non-perturbative D-brane dynamics, which is still lacking.  
Nevertheless, we believe that the scenario for supersymmetry obstruction
presented here is of sufficient interest to motivate 
further detailed studies of such issues. 

\section*{Acknowledgments} 

One of us (N.E.M.) wishes to thank K. Tamvakis for enlightening
discussions on the r\^ole of conical singularities in supersymmetry
obstruction.  A.C.--S. takes pleasure in thanking Subir Sarkar for
helpful discussions.  The work of N.E.M.  is partially supported by a
P.P.A.R.C. (U.K.)  Advanced Fellowship. A.C.--S. is grateful to
P.P.A.R.C. (U.K.) for a research studentship (number 96314661).

\end{document}